\documentclass[14 points]{article}
\usepackage{amsmath, amssymb, amsthm}
\usepackage{geometry}
\usepackage{graphicx}
\usepackage{hyperref}
\usepackage{subcaption}
\usepackage{setspace}
\usepackage{authblk}

\begin{document}

\title{A Novel Framework for Comparing Combination Therapy Outcomes Using Mechanistic Graph Models}

\author[1]{Dr. Dipayan Sengupta, MD (Dermatology)}
\author[2]{Dr. Saumya Panda, MD (Dermatology)}

\affil[1]{Consultant Dermatologist, Charnock Hospital, Kolkata, India}
\affil[2]{Professor and Head, Department of Dermatology, Jagannath Gupta Institute of Medical Sciences and Hospital, Kolkata, India}

\date{}

\maketitle

\vspace{1cm}
\noindent\textbf{Manuscript Word Count:} 6524 words\\

\noindent\textbf{Running Head:} Efficacy comparison framework using graph model\\

\vspace{1cm}

\section*{Abstract}
\textbf{Background}: Predicting the efficacy of combination therapies is a critical challenge in clinical decision-making, particularly for diseases requiring multi-drug regimens. Traditional evidence synthesis methods, such as component network meta-analysis (cNMA), often face parameter explosion and limited interpretability, especially when modeling interaction effects between components.

\textbf{Objective}: This article introduces a general Efficacy Comparison Framework (ECF), a mechanistically grounded system for predicting combination therapy outcomes. ECF integrates biological pathway-based abstractions with expert knowledge, optimized with quasi-rules derived from  clinical trial data to overcome the limitations of traditional methods.

\textbf{Methods}: ECF employs a disease pathogenesis graph to encode domain knowledge, reducing the parameter space through mechanistic functions and sparse network structures. Optimization may be performed using a loss function inspired by the Thurstone-Mosteller model, focusing on pairwise regimen comparisons. A pilot study was conducted for acne vulgaris to evaluate ECF’s ability in both tested and untested comparisons.

\textbf{Results}: In the acne vulgaris case study, the ECF-based model achieved \textbf{76\% accuracy} in predicting both tested and untested regimen outcomes, demonstrating statistically comparable performance across clinical trial data and expert dermatologist consensus (\textbf{p = 0.977}). The agreement between ECF and expert predictions was within the range of \textbf{inter-expert agreement}, supporting the model’s potential as a dynamic system that can mimic clinicians' intuition.

\textbf{Discussion}: ECF aligns with recent advancements in network science and synergy prediction, leveraging principles of complementary targeting and biological plausibility. Its use of disease pathogenesis graphs offers a more interpretable and scalable alternative to existing models reliant on chemical similarity or protein-protein interaction (PPI) topology.

\textbf{Conclusion}: ECF represents a significant advancement in evidence synthesis for combination therapy, providing a biologically grounded, scalable and interpretable framework. It holds promise for guiding personalized medicine, developing treatment guidelines, and advancing systems pharmacology, with future directions focusing on multi-omics integration, toxicity prediction, and validation of untested regimen predictions.

\section*{1. Introduction}

Comparing the efficacy of combination therapies is a pressing challenge in modern clinical decision-making, particularly in complex diseases where clinicians frequently employ \textit{untested} therapeutic regimens [\textit{1}]. Being able to compare and select optimal combinations is critical, irrespective of whether the knowledge originates from \textbf{clinical studies} or \textbf{broader medical expertise}. In the \textbf{classical evidence synthesis} paradigm, \textbf{network meta-analysis (NMA)} and its extension, \textbf{component network meta-analysis (cNMA)}, are fundamentally suited to this objective [\textit{2}], but they face fundamental limitations when it comes to \textbf{interaction} terms in combination therapy.

---

\subsection*{1.1 Background: NMA and cNMA for Combination Therapy}
Traditional \textbf{NMA} enables indirect comparisons among multiple interventions by pooling data across trials, often focusing on single interventions or pairwise comparisons [\textit{3}]. \textbf{cNMA} generalizes this idea to \textbf{multi-component} treatments, breaking a combination therapy into individual drugs (components) and modeling their separate contributions:
\[
\eta_T 
\;=\; 
\beta_0 
\;+\; 
\sum_{i \in T} \beta_i,
\]
where \(T\subseteq\{1,\ldots,n\}\) is a subset of \(n\) possible drug components, \(\beta_i\) is the main effect of component \(i\), and \(\beta_0\) is a baseline effect [\textit{4}]. This \textbf{additive cNMA}, however, ignores \textbf{synergy} or \textbf{antagonism}, a critical aspect of combination regimens that clinicians already exploit by combining drugs with differing mechanisms of action (MOA).

To address synergy, \textbf{interaction cNMA} introduces pairwise \textbf{interaction} terms \(\theta_{ij}\):
\[
\eta_T 
\;=\;
\beta_0 
\;+\; 
\sum_{i \in T} \beta_i
\;+\;
\sum_{\substack{i, j \in T,\,i < j}} \theta_{ij},
\]
where \(\theta_{ij}\) captures the synergy (or antagonism) between components \(i\) and \(j\). This is more aligned with clinical intuition—physicians prefer combining interventions with complementary MOAs precisely because of these \textbf{interaction} effects (also recently shown by a human protein-protein interactome based approach [\textit{5}]). But implementing interaction cNMA at scale introduces \textbf{major challenges} of dimensionality.

\subsection*{1.2 Two Approaches to Comparing Untested Regimens and Their Limits}

Clinicians (and researchers) may conceptualize the comparison of untested multi-drug regimens in \textbf{two} broad ways:
\begin{enumerate}
    \item \textbf{Estimating a Mean Efficacy} (or outcome) for each regimen and comparing those values, exemplified by \textbf{interaction cNMA}.
    \item \textbf{Directly contrasting} regimens in a \textbf{Bradley-Terry type} (BT) framework [\textit{6}], assigning probabilities \(P(T \succ S)\) that regimen \(T\) outperforms \(S\) in a head-to-head sense.
\end{enumerate}

While these two paradigms differ mathematically, \textit{both} encounter \textbf{combinatorial burdens} when the set of potential combinations grows exponentially with the number of drug components.

\subsubsection*{Approach A: Interaction cNMA for Mean Efficacy}

\textbf{Interaction cNMA} remains the most natural extension of cNMA to handle synergy. However, a \textbf{Bayesian} version typically requires setting priors on the \(\theta_{ij}\) parameters, each representing synergy or antagonism between drugs \(i\) and \(j\). For local pairwise comparisons—say analyzing a single triple regimen—this might be tractable. But as soon as one aims to compare \textit{all} or \textit{many} multi-drug regimens of size 3, 4, or 5, the number of interaction terms \(\binom{n}{2}\) grows \textbf{quadratically}, leading to:
\begin{itemize}
    \item \textbf{Combinatorial Prior Specification}: If we attempt to incorporate domain knowledge for each synergy term, we face an overwhelming burden to set (or learn) \(\binom{n}{2}\) priors.
    \item \textbf{Sparse Data}: Most synergy terms might never be directly estimated if trials cover only a small fraction of possible combinations.
    \item \textbf{High Variance} for large multi-drug predictions if synergy parameters remain underinformed.
\end{itemize}

Hence, while theoretically powerful, \textbf{interaction cNMA} is difficult to operationalize at large scale.

\subsubsection*{Approach B: Direct Contrastive (Bradley-Terry Type) Models}

Alternatively, we might \textbf{compare} multi-drug regimens \textit{directly} rather than assigning each a numeric efficacy. A \textbf{Bradley-Terry (BT)} approach would place each regimen \(R\) in a latent scale \(u_R\) and model pairwise comparisons as:
\[
P(R \succ S) = \frac{\exp(u_R)}{\exp(u_R) + \exp(u_S)}.
\]
If we have \(N\) total regimens, a naive bounding argument suggests we need on the order of \(N \log N\) independent comparisons as the lower bound [\textit{7}], to stably rank or compare all \(N\) items. Given \(n\) drug components, there are \(N = 2^n - 1\) nonempty combinations. Then the naive requirement grows as \((2^n)\log(2^n) \approx n \cdot 2^n\)—again infeasible for typical clinical datasets where only a handful of combinations are compared in literature.

In sum, \textbf{both} interaction cNMA (estimating a separate synergy parameter for each pair) and direct BT-style models (requiring \(\mathcal{O}(N\log N)\) comparisons) face \textbf{combinatorial explosions} in large multi-drug scenarios. The question then becomes: how do we incorporate \textbf{domain knowledge} to reduce dimensional complexity?

\subsection*{1.3 Mechanistic Hypothesis: Lower-Level Abstractions}

In order to \textbf{reduce} the dimension of synergy parameters while simultaneously retaining the ability to model interactions, we assume that emergence of complex interaction can be traced back to a \textbf{lower-level biological} or mechanistic layer, which is drastically smaller in dimension than enumerating all drug pairs [\textit{8}]. This idea can be applied in both the cNMA and BT contexts:

\begin{enumerate}
    \item \textbf{Mechanistic decomposition of interaction cNMA}: Suppose we define \(m \ll n\) \textbf{mechanistic variables} (e.g., pathways, drug targets). Each drug \(i\) exerts an effect via \(\delta_i = g_i(M_1,\dots,M_m)\). Pairwise synergy emerges automatically if two drugs share or interact through overlapping mechanistic variables. The parameter count is \(\mathcal{O}(n \cdot m)\) rather than \(\binom{n}{2}\).
    \item \textbf{\(d\)-Dimensional Embedding for BT model}: If each component can be placed in a \textbf{\(d\)-dimensional} latent space, the number of pairwise comparisons needed to rank them shrinks from \(\mathcal{O}(N\log N)\) to \(\mathcal{O}(dn \log n)\) [\textit{9}] or approximately \(\mathcal{O}(d.n)\) . The intuition is that if the data truly lie in a small \(d\)-dimensional manifold with \(n\) truly independent players, we do not require a complete set of \(\binom{N}{2}\) matchups. This requirement does not change even when combinations are used as linear aggregation preserves the low-rank structure of the interaction matrix, allowing the sample complexity to inherit guarantees from low-rank matrix completion [\textit{10}]
\end{enumerate}

Conceptually, the mechanistic variables (\(m)\) and dimensions (\(d\)) in the embedding space are the same thing and imagined as the lower level abstraction of individual treatment components effect. However, it is worth noting that \(d\)-dimensional embedding would mostly imagine combination regimens as linear combination of components [\textit{11}] similar to additive CNMA (albeit in a lower dimension). Though other approaches exist [\textit{12}], they are not biologically intuitive as well.

\subsection*{1.4 Few Core Intuitions for Further Dimensional Control}
While adopting \textbf{mechanistic variables} addresses the large-scale dimensional issue, few additional refinements can drastically cut the parameter space and reduce posterior uncertainty while incorporating domain knowledge in the process:

\subsubsection*{1.4.1 Intuition 1: Mechanistic Graphs}
In many disease contexts, variables \(\{M_1,\dots,M_m\}\) form a \textbf{sparse network} \(\mathcal{G}=(V,E)\). Each node \(v\in V\) corresponds to a subset of \(\{M_1,\dots,M_m\}\). Here, interaction terms are drastically reduced because each intervention interacts with only a few nodes instead of all other interventions. Domain knowledge can be more naturally incorporated while building the graph, specifying the valid node-intervention interactions as well as setting their initial values prior to optimization.
 The resulting “graph-limited synergy” is far below \(\binom{n}{2}\) in typical sparse systems, reflecting a biologically motivated curation of plausible interactions. 

\subsubsection*{1.4.2 Intuition 2: Effect-Based Synergy Aggregators}
Beyond restricting which drug pairs can interact (indirectly), we can also incorporate domain knowledge by controlling the functions that govern the cumulative effect of interventions and parenteral influence on a specific node at a specific time step. Intuitively, depending on the available knowledge, we can choose appropriate \textbf{pharmacologically grounded aggregator functions} (instead of linear addition ). Classical effect-based combination formulas—\textbf{Bliss Independence}, \textbf{Highest Single Agent (HSA)}, or \textbf{simple additivity}—encode baseline synergy/antagonism without enumerating free coefficients for each pair [\textit{13}]. These models do not need concentration-response curves (unlike dose-effect based models such as Loewe’s additivity or quantitative systems pharmacology(QSP)-based rate laws), which are often unavailable in standard clinical trials. Any additional deviation (a “true” synergy parameter) is introduced \textbf{only} where strong mechanistic evidence suggests a unique phenomenon (e.g., direct receptor competition). For instance, if two drugs \(i\) and \(j\) converge on a node with a synergy aggregator \(f(\alpha_i,\alpha_j)\), we might have:
\[
\eta_{ij} 
\;=\; 
f(\alpha_i,\alpha_j) 
\;+\;
\Delta_{ij},
\]
where \(\Delta_{ij}\) is assigned a prior of near zero unless domain knowledge justifies significant synergy. This approach drastically reduces the synergy dimension and integrates standard pharmacological models into the framework making it more biologically plausible.

\subsubsection*{1.4.3 Intuition 3: optimum complexity of the graph can be determined beforehand}

The effective complexity of a mechanistic graph can be inferred from the number of independent pairwise comparisons \(S\) available in the data. As seen in section \textbf{1.3}, for pairwise comparison, the parameter space scales as \(\mathcal{O}(n \cdot m)\), where \(n\) is the number of treatment components and \(m\) is the number of mechanistic variables. 

However, a more fundamental derivation of the lower bound on the number of observations required to estimate \( N \) unknown parameters arises from information-theoretic principles. Given a system with \( N \) parameters, each taking values from a discrete set of size \( M \), the total number of possible configurations is \( |\Theta| = M^P \), and the total entropy of the parameter space is given by:

\[
H(\Theta) = N \log M.
\]

In the worst-case scenario where parameters are maximally uncertain, a uniform prior is assumed, meaning each parameter takes \( M = O(N) \) possible states. Substituting this into the entropy formula, we obtain:

\[
H(\Theta) = N \log N.
\]

This entropy quantifies the number of bits required to uniquely specify the parameter space before any data is observed [\textit{14}]. Estimating \( \Theta \) from observations \( Y \) requires reducing this entropy via mutual information \( I(\Theta; Y) \), which quantifies how much information the data provides about the parameters. From Fano’s inequality [\textit{14}], the probability of error in estimating \( \Theta \) satisfies:

\[
P_e \geq \frac{H(\Theta) - I(\Theta; Y) - 1}{\log |\mathcal{A}|}.
\]

For small error probability (\( P_e < \frac{1}{2} \)), we require that the mutual information be sufficiently large:

\[
I(\Theta; Y) \geq H(\Theta) - 1.
\]

Since each independent observation provides at most \( O(1) \) bit of information, the total mutual information from \( S \) observations is bounded by:

\[
I(\Theta; Y) \leq S \cdot I(\Theta; y_{ik}),
\]

where \( I(\Theta; y_{ik}) = O(1) \) in the binary outcome case. Substituting this into the Fano bound and solving for \( S \), we obtain the fundamental scaling law:

\[
S = \Omega\left(\frac{H(\Theta)}{I(\Theta; y_{ik})}\right) = \Omega(N \log N).
\]

For a fixed \( S \), we can solve this for \( N \) using the Lambert function. However, as this derivation is based on the assumption of maximum uncertainty (i.e., no prior knowledge), in practice, this can be slightly relaxed as domain knowledge is supposed to be integrated at various levels. However, the exact parameter size reduction is difficult to estimate (should be proportional to the reduction of Kullback–Leibler divergence) as it is difficult to quantify the impact of domain knowledge in this setting.
Additionally, in sparse graphs, not all \(m\) interact with every \(n\), reducing the effective parameter space to \(\mathcal{O}(n \cdot d)\), where \(d\) is the average degree (i.e., the average number of mechanistic variables interacting with each component). 
By leveraging \(S\) to constrain \(d\) and \(m\), the graph complexity can be tailored to match the available evidence, preventing over-fitting while maintaining sufficient flexibility to encode mechanistic interactions.

\subsection*{1.5 Toward the new framework}

These \textbf{mechanistic} underpinnings—(1) an abstraction to \textbf{lower-level biological variables} and (2) \textbf{effect-based aggregator} functions—set the stage for a \textbf{novel framework} which would leverage a \textbf{graph-based} representation of disease, with each drug mapping onto certain nodes (mechanisms) via weight vectors. Combinations yield synergy implicitly at each node aggregator, circumventing the need to enumerate synergy parameters for every pair/trio of drugs. Moreover, this structure readily accommodates domain knowledge:

In the \textbf{next section}, we formally introduce the new framework with its discrete-time mechanistic structure, aggregator definitions, and implementation details, illustrating how these ideas deliver a \textbf{scalable}, \textbf{clinically relevant} approach to multi-drug evidence synthesis that retains the essence of \textbf{interaction cNMA} while avoiding its parameter explosion.

\section*{2. Method: Formal description of our framework}

We hereby introduce a scalable \&\ flexible framework designed to incorporate domain knowledge into the available clinical evidence while avoiding parameter explosion or over-fitting. The goal of this framework is to propose some general strategies to build graph-based disease specific models which will be able to compare any combination regimen irrespective of whether it was previously tested or not. For the ease of description, we will call this \textbf{Efficacy Comparison Framework (ECF).}

Below, we are describing the general form of ECF with possible approaches. The core idea is to incorporate mechanistic knowledge at a lower level of abstraction through a graphical influence diagram with controlled complexity. But design of each components (parameter or function selection or optimization techniques) can vary depending on the research objective.

\subsection*{2.1 Graph-Based Mechanistic Core of ECF}

ECF starts from a \textbf{directed graph} $\mathcal{G} = (V, E)$, where:
\begin{itemize}
    \item \textbf{Nodes} ($v \in V$) represent mechanistic variables, such as biological pathways, drug targets, or intermediate processes relevant to disease progression.
    \item \textbf{Edges} ($e \in E$) represent plausible interactions between these variables, encoding biological dependencies or causal links.
\end{itemize}

\subsubsection*{Drug--Node Mapping}
Each drug $k$ acts on a subset of nodes in $\mathcal{G}$. Let $w_{k \to v}$ represent the weight of the effect of drug $k$ on node $v$. For $n$ drugs and $m$ nodes, we define a matrix of drug-to-node weights:
\[
W = [w_{k \to v}], \quad k = 1, \dots, n, \; v = 1, \dots, m,
\]
where:
\begin{itemize}
    \item $w_{k \to v} \geq 0$ if drug $k$ targets node $v$, and $w_{k \to v} = 0$ otherwise.
    \item $\sum_{v=1}^m w_{k \to v} = 1$, ensuring each drug's total effect is distributed across its target nodes.
\end{itemize}

This drug-to-node mapping reduces parameter complexity by focusing on \textbf{mechanistic variables} rather than directly modeling pairwise drug interactions.

While constructing the core graph, it may be tempting to increase complexity by incorporating finer-grained biological data, such as gene- or protein-level interactions. However, we caution that the graph complexity should be chosen according to the available dataset for optimization, as established in Section 1.4.3. Overly complex graphs may require more parameters than the data can support, leading to overfitting and unreliable predictions. For practical implementation, \(d\) (the average degree) and \(m\) (the total number of nodes) should align with the available comparisons \(S\) to ensure that the graph structure balances biological fidelity and computational feasibility. This alignment ensures that the graph complexity reflects the true dimensionality supported by the dataset.

\subsection*{2.2 Discrete-Time Node Updates and Synergy Modeling}

ECF should commonly operate in discrete time steps $\Delta t$, corresponding to the intervals commonly reported in clinical trials (e.g., baseline, 4 weeks, 8 weeks). At each time step, the value $N_v(\Delta t)$ of node $v$ evolves based on:
\begin{enumerate}
    \item \textbf{The node’s previous state} ($N_v(\Delta t - 1)$).
    \item \textbf{Cumulative drug effects} ($C_v(\Delta t)$).
    \item \textbf{Influences from parent nodes} ($u$ connected to $v$ via $e \in E$).
\end{enumerate}

\subsubsection*{Node-Level Cumulative Drug Effect}
The cumulative effect of all drugs $r_v(\Delta t)$ targeting node $v$ is computed using an \textbf{aggregator function} $f_v$:
\[
C_v(\Delta t) = f_v\bigl(\alpha_{1 \to v}, \alpha_{2 \to v}, \dots, \alpha_{r \to v}\bigr),
\]
where:
\begin{itemize}
    \item $\alpha_{k \to v} = w_{k \to v} \cdot E_k(\Delta t)$ is the scaled effect of drug $k$ on node $v$ at time $\Delta t$.
    \item $E_k(\Delta t)$ is the efficacy of drug $k$ as determined by trial data or priors.
\end{itemize}

Based on pharmacological principle, common choices for $f_v$ may include:
\begin{enumerate}
    \item \textbf{Bliss Independence} (for independent pathways with same end target):
    \[
    f_v(\alpha_1, \alpha_2) = \alpha_1 + \alpha_2 - \alpha_1 \alpha_2.
    \]
    This is extended for $r > 2$ drugs as:
    \[
    f_v(\alpha_1, \alpha_2, \dots, \alpha_r) = \sum_{i=1}^r \alpha_i - \prod_{i=1}^r (1 - \alpha_i).
    \]
    \item \textbf{Highest Single Agent (HSA)} (competitive binding on same pathway):
    \[
    f_v(\alpha_1, \alpha_2, \dots, \alpha_r) = \max(\alpha_1, \alpha_2, \dots, \alpha_r).
    \]
    \item \textbf{Additive Effect} (independent pathways without saturation acting on different end target):
    \[
    f_v(\alpha_1, \alpha_2, \dots, \alpha_r) = \sum_{i=1}^r \alpha_i.
    \]
Domain knowledge should be used to set aggregator functions in the right context. 
    
\end{enumerate}

\subsubsection*{Node Evolution Over Time}
The state of node $v$ at time $\Delta t+1$ depends on its previous state, cumulative drug effects, and influences from parent nodes. A discrete-time update equation for node $v$ is:
\[
N_v(\Delta t+1) = g_v\bigl(N_v(\Delta t), C_v(\Delta t), \{\Delta N_u(\Delta t)\}_{(u \to v) \in E}\bigr),
\]
where:
\begin{itemize}
    \item $g_v$ is a nonlinear function modeling node behavior.
    \item $\Delta N_u(\Delta t) = N_u(\Delta t) - N_u(\Delta t-1)$ is the change in parent node $u$'s value.
\end{itemize}

This formulation avoids the need for continuous data or ODE modeling, aligning ECF with the data granularity available in clinical trials.

Also it is important to note that this operations can be viewed as a specialized representation of a dynamic Bayesian network whose general form for a node value $X_i^t$ can be derived as:
\[
P(X_i^t \mid \text{Parents}(X_i^t), \text{Parents}(X_i^{t-1}))
\]
[\textit{15}]

Here the conditional probability model is being represented with appropriate biological aggregator functions similar to structural causal modeling. However, at its core, it makes the same assumption of conditional independence.

\subsubsection*{1. Conditional Independence}
ECF must assume that the state of each node $v \in V$ is conditionally independent of all other nodes given its parent nodes in $\mathcal{G}$. Formally:
\[
p(N_v \mid N_u, u \notin \text{Pa}(v)) = p(N_v \mid \text{Pa}(v)),
\]
where $\text{Pa}(v)$ represents the parent nodes of $v$.

\subsubsection*{2. Markov Property}
The temporal evolution of each node is governed by a Markov process. That is, the value of a node at time step $t+1$ depends only on its value at $t$ and its inputs at $t$, ensuring that:
\[
p(N_v(t+1) \mid N_v(0), N_v(1), \dots, N_v(t)) = p(N_v(t+1) \mid N_v(t)).
\]

\subsection*{2.3 Parameter Space and Scalability}

The parameter space in ECF grows linearly with $n$ (number of drugs) and $m$ (number of nodes) rather than quadratically or exponentially. The total number of parameters is approximately:
\[
N_{\text{ECF}} = n \cdot m + |E| + \text{parameters in } f_v.
\]
where \(|E|\) is number of extra parameters due to graph structure. This is significantly smaller than the parameter count in interaction cNMA:
\[
N_{\text{cNMA}} = n + \binom{n}{2},
\]
where $\binom{n}{2}$ arises from the combinatorial growth of pairwise interaction terms (may be more than 2 for multi-way interaction).

The reduction is due to:
\begin{enumerate}
    \item \textbf{Node-Level Aggregation}: Synergy is encoded through $f_v$, not enumerated for all drug pairs.
    \item \textbf{Sparse Graphs}: The disease graph $\mathcal{G}$ limits interactions to plausible mechanistic relationships, further reducing parameters.
\end{enumerate}

\subsection*{2.4 Optimization Approaches for Parameter Fitting}

The optimization of our framework is inherently focused on predicting regimen superiority in the presence of non-linear systemic modularity. As seen in other domains [\textit{16}], we think that a contrastive loss function would be most appropriate. While many optimization approaches could theoretically be employed, we propose a system inspired by the \textbf{Thurstone-Mosteller (T-M) Model} [\textit{17}] due to its conceptual similarity to our objective. The T-M model's probabilistic framework, which uses statistical distributions derived from clinical trial data to calculate win probabilities between regimens, provides a natural foundation for this task. However, the approach we describe is largely based on the authors' conceptual understanding and we understand that many different optimization approaches are possible.

\subsubsection*{2.4.1 Probabilistic Foundation for Regimen Comparison}

The \textbf{win probability} of regimen $T_A$ being superior to $T_B$ can be calculated using the Thurstone-Mosteller model, which assumes treatment outcomes follow normal distributions:
\[
P(T_A > T_B) = \Phi\left(\frac{\mu_A - \mu_B}{\sqrt{\sigma_A^2 + \sigma_B^2}}\right),
\]
where:
\begin{itemize}
    \item $\mu_A, \mu_B$: Expected outcomes (mean effect sizes) for regimens $T_A$ and $T_B$,
    \item $\sigma_A^2, \sigma_B^2$: Variances of outcomes derived from trial data,
    \item $\Phi$: Cumulative distribution function (CDF) of the standard normal distribution.
\end{itemize}

This probabilistic approach aligns naturally with clinical trial data, where mean differences ($\mu_A - \mu_B$) and pooled variances ($\sigma_A^2 + \sigma_B^2$) are readily available or calculable.

\subsubsection*{2.4.2 Objective Function for Optimization}

While many objective functions are possible, here we describe a possible frequentist approach. Let's have an error function $\mathcal{E}$ that penalizes mismatches in \textbf{directionality} (regimen superiority predictions) between ECF outputs and observed regimen comparison. Let:
\begin{itemize}
    \item $\Delta_{\text{ECF}}(T_A, T_B) = W_A - W_B$: The ECF-predicted difference in cumulative weight reductions in the final output node for regimens $T_A$ and $T_B$,
    \item $\Delta_{\text{meta}}(T_A, T_B) = \mu_A - \mu_B$: The observed effect size difference from meta-analysis (if multiple studies are available for that comparison),
    \item $P(T_A > T_B)$: The win probability derived from the T-M model.
\end{itemize}

The optimization minimizes the following error function:
\[
\mathcal{E} = \sum_{T_A, T_B \in \mathcal{C}} P(T_A > T_B) \cdot \mathbb{I} \left(\text{sgn}(\Delta_{\text{ECF}}(T_A, T_B)) \neq \text{sgn}(\Delta_{\text{meta}}(T_A, T_B))\right),
\]
where:
\begin{itemize}
    \item $\mathcal{C}$: Set of all pairwise comparisons,
    \item $\mathbb{I}(\cdot)$: Indicator function, evaluating to 1 if ECF's directional prediction disagrees with the observed clinical evidence.
\end{itemize}

The inclusion of $P(T_A > T_B)$ ensures that comparisons with stronger clinical evidence (lower uncertainty) contribute more heavily to the optimization process.

\subsubsection*{2.4.3 Flexibility in Win Probability}

While the T-M model provides a literature derived win probability, flexibility is introduced to account for real-world factors such as:
\begin{itemize}
    \item \textbf{Bias Adjustment}: Win probabilities can be adjusted for study-level biases (e.g., industry funding, poor methodology).
    \item \textbf{Expert Overrides}: Optionally experts can modify probabilities based on external knowledge, such as observational data or specific patient subgroups. These flexibilities are expected to make the model more robust specially when high quality data is limited.
\end{itemize}

These adjustments can be incorporated post hoc, ensuring that the system remains adaptable to varying clinical contexts while preserving mathematical rigor. Few possible extensions have been proposed here [\textit{18}] for bias adjustment.

Being a flexible framework, we acknowledge that there is significant room for refinement or alternative implementations based on model objectives, resources or other factors. For example, one may use a maximum likelihood estimation for each parameter, replacing win probability with standard effect-size measurement, or even a Bayesian approach for inference is possible (though likely to be computationally intensive).

\section*{3. Experiment}

We conducted a simple proof-of-concept pilot study to build a model using ECF for acne vulgaris (selected due to its wide range of treatment options and the frequent use of combination regimens in clinical practice). It is however, to be noted that, ECF is a general framework, not a rigid statistical method. So, multiple implementations for the same condition is possible with different assumptions, graph complexity, optimization technique etc depending on available data, computational resource and research objective. 

\subsection*{3.1 Building the Basic Framework for Acne Vulgaris}

\textbf{Graph Construction}  

A model based on ECF specifically designed for acne vulgaris was developed, focusing on common medical interventions while excluding procedural therapies. The model's foundation was a graph representation of acne pathogenesis, capturing the mechanisms of various interventions. Only nodes and connections with significant pathogenetic and therapeutic implications were included for simplicity (\textbf{Figure 1a}).

\textbf{Intervention Selection}  

We conducted a systematic literature search, including guidelines published in the last seven years [\textit{19,20,21,22,23,24,25,26}] up to October 2023, meeting the AGREE II criteria [\textit{27}] for rigor of development. First- and second-line treatments backed by Level A/B or Grade A/B evidence were considered. Interventions explicitly discouraged in guidelines were excluded, but regional exclusions (e.g., dapsone gel unavailability in Singapore) were not considered. However, treatments available only on specific regions, such as Keigairengyoto in Japan, were also omitted. 

Absence of recommendation in a guideline was not a reason for exclusion unless another guideline explicitly advised against its use. Maintenance and adjuvant therapies were excluded. The finalized intervention list (\textbf{Table 1}) avoided sub-classification, such as distinguishing between oral contraceptive formulations.

\begin{table}[ht]
\centering
\begin{tabular}{|l|l|}
\hline
\textbf{Group}          & \textbf{Interventions}                                                                                   \\ \hline
\textbf{Topical Retinoids} & Adapalene, Tretinoin 0.025\%, Tretinoin 0.05\%, Tazarotene, Isotretinoin gel                            \\ \hline
\textbf{Topical Antibiotics} & Clindamycin, Benzoyl peroxide, Erythromycin, Dapsone, Ozenoxacin, Nadifloxacin                       \\ \hline
\textbf{Other Topical}   & Azelaic acid                                                                                              \\ \hline
\textbf{Oral Retinoids}  & Isotretinoin (standard dose), Isotretinoin (low dose)                                                     \\ \hline
\textbf{Oral Antibiotics} & Doxycycline, Azithromycin (pulse dosage), Lymecycline, Roxithromycin                                    \\ \hline
\textbf{Hormonal Therapies} & Metformin, Spironolactone, Oral contraceptive                                                           \\ \hline
\end{tabular}
\caption{List of selected Interventions}
\label{tab:interventions}
\end{table}

Each intervention was linked to at least one node in the graph, representing its mechanism of action. \textbf{Figure 1b} illustrates an example where a node is influenced by both its parent nodes and associated interventions. Parameters were initialized using expert knowledge and constrained within plausible biological ranges.

\begin{figure}[htp]
    \centering
    \begin{subfigure}[b]{0.45\textwidth}
        \centering
        \includegraphics[width=\textwidth]{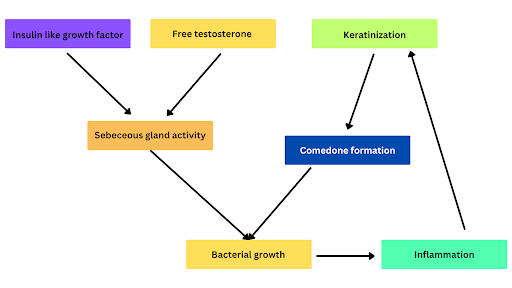}
        \caption{All nodes with connecting edges in our model. The direction of arrow indicates the direction from parent node to the child node. The graph is not acyclic because the event goes from ‘inflammation’ to ‘keratinization’ \& vice versa. The interventions for each nodes have been mentioned in Table 3 }
        \label{fig:1a}
    \end{subfigure}
    \hfill
    \begin{subfigure}[b]{0.45\textwidth}
        \centering
        \includegraphics[width=\textwidth]{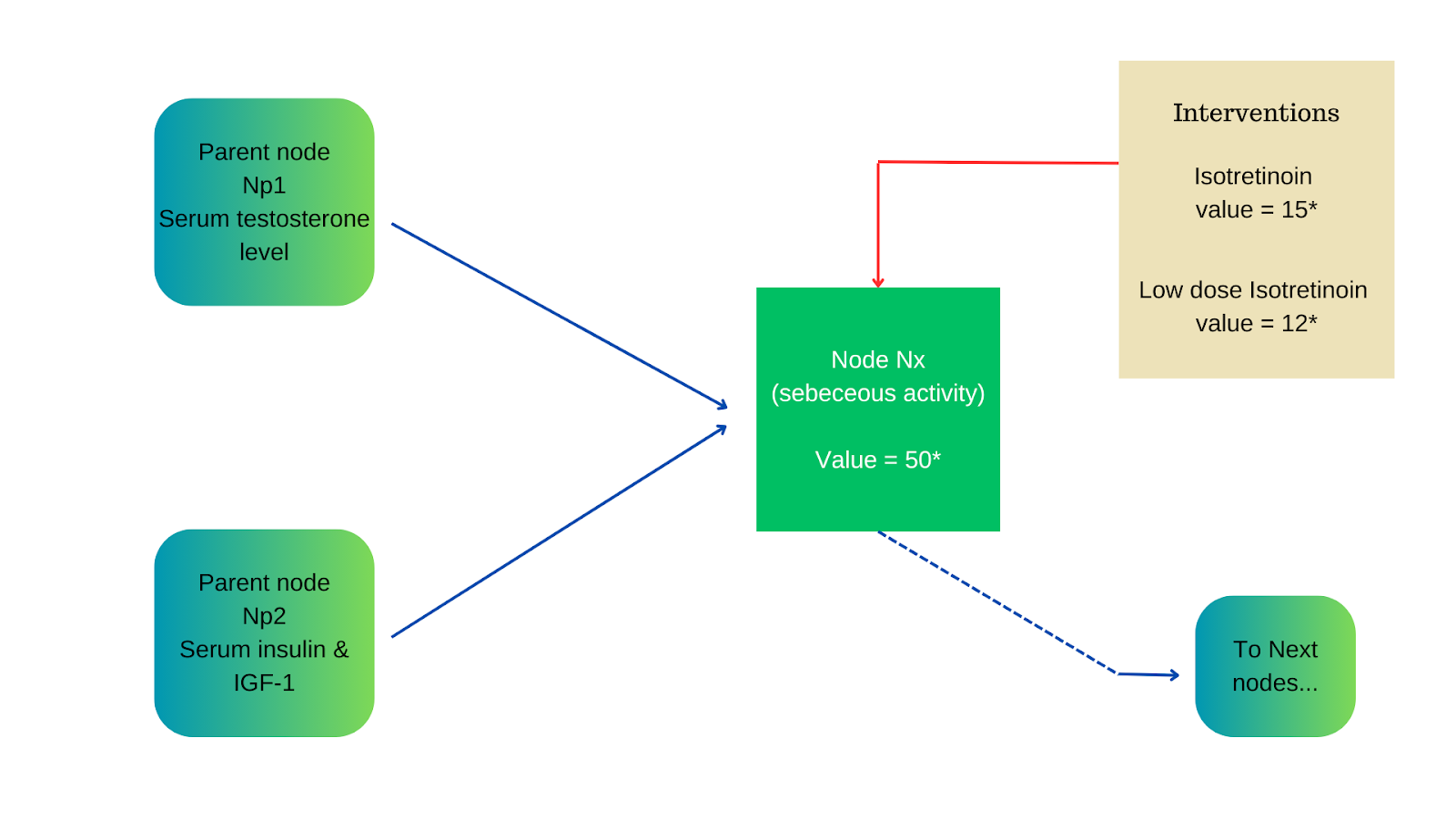}
        \caption{Outline of the proposed model's graphical structure, taking sebaceous gland activity as an example node. It is influenced by the parent nodes while being inhibited by the interventions acting on it such as isotretinoin. The node’s initial value and the weights of the interventions are to be optimized with quasi-rules taken from the literature (see text for details).}
        \label{fig:1b}
    \end{subfigure}
    \caption{}
    \label{fig:main}
\end{figure}

\textbf{Setting Up Functions}  

The cumulative effect functions ($ g_v(\cdot) $ , $ C_v(\cdot) $ etc) in this model were adapted as follows:

\begin{enumerate}
    \item \textbf{Cumulative Effects}: The Highest Single Agent (HSA) model was used for interventions targeting the same pathway [\textit{28}], while Bliss Independence (BI) was employed for those acting on distinct pathways at the same node [\textit{29}]. No additional true interaction between treatment components was assumed.

    \item \textbf{Node Aggregation}: BI was also applied for combining the influences of parent nodes, in alignment with the conditional independence assumption of Bayesian networks and structural causal models.
\end{enumerate}

\textbf{Training and Optimization}  

The model was optimized using quasi-rules derived from the literature, formatted in a standard evidence-based PICO structure. For example, a rule might state: "Topical adapalene is more efficacious than topical tretinoin (0.025\%) at 8 weeks for inflammatory acne vulgaris." Publicly accessible literature from Google and Google Scholar was used without reliance on a specific database. Each rule was assigned an importance score indicative of the win probability of the superior regimen, as outlined in the method section (refer to \textbf{Table 2}). 

To accommodate qualitative factors such as bias and lower-quality evidence, precise win probability calculations were not implemented in this version. Future iterations will require a robust, standardized framework to improve the precision of these importance scores.

\begin{table}[ht]
\centering
\begin{tabular}{|l|l|l|l|l|l|l|}
\hline
\textbf{Group A}         & \textbf{Group B}         & \textbf{Winner} & \textbf{Type of acne} & \textbf{Timespan} & \textbf{Acne Grade} & \textbf{Importance} \\ \hline
doxy                     & azi, clinda              & b                & both                   & t3                 & 6                   & 8                   \\ \hline
doxy, dapsone             & doxy, bpo                & equal            & inflammatory           & t3                 & 5                   & 6                   \\ \hline
doxy, bpo                 & doxy, dapsone            & a                & comedonal              & t2                 & 5                   & 6                   \\ \hline
doxy, ada                 & azi, ada                 & a                & both                   & t3                 & 3                   & 7                   \\ \hline
azi                       & ada                      & b                & comedonal              & t3                 & 3                   & 10                  \\ \hline
azi, ada                  & azi                      & a                & inflammatory           & t3                 & 3                   & 10                  \\ \hline
azi, ada                  & ada                      & a                & inflammatory           & t3                 & 3                   & 10                  \\ \hline
doxy                      & clinda                   & equal            & both                   & t3                 & 6                   & 8                   \\ \hline
iso                       & doxy, bpo, ada           & a                & both                   & t4                 & 7                   & 8                   \\ \hline
\end{tabular}
\caption{A set of quasi-rules used for building the model}
\label{tab:quasi_rules}
\end{table}

\textbf{Graph Complexity Determination}:
In the case study of acne vulgaris, we constructed an optimization dataset consisting of approximately 110 quasi-rules derived from clinical guidelines and evidence, covering around 20 interventions (treatment components). Based on the relationship \( N \log N \approx S \), we estimated the total number of reliably identifiable interactions \( N \) by solving the equation:

\[
N = \exp(W(S)),
\]

where \( W(S) \) is the Lambert function. Using this approach, we found \( N \approx 31.8 \) as the most accurate estimate. Approximating the number of mechanistic interactions per intervention as \( m = N/n \), we obtained:

\[
m \approx 1.59.
\]

Consequently, we opted for a graph with 7 nodes, while each intervention interacts with only one or two mechanistic components ensuring that the mechanistic pathways captured the core biological processes of acne pathogenesis while maintaining computational feasibility, as discussed in \textbf{Section 1.4.3}.

\subsection*{3.2 Optimization of our model}

A hierarchical optimization strategy (block coordinate descent) was implemented, as described in Section 2.

\textbf{Stage 1: Intra-Group Optimization}:  
Interventions within the same category (e.g., topical retinoids or systemic antibiotics) were optimized against one another using group-specific rules. For example, comparisons between two topical retinoids or between two systemic antibiotics were performed separately.

\textbf{Stage 2: Inter-Group Optimization} :
In the second stage, comparisons were extended across different categories, such as between topical retinoids and systemic antibiotics. This stage ensured that the relative positions of interventions across categories were consistent with the defined rules. Convergence was achieved when the error function \(\epsilon\) stabilized (\( \Delta \)\( \epsilon \) < 1\% over 50 iterations). No hyperparameter tuning was required, as the learning rate $\eta$ was fixed at 0.01.

This two-step approach reduced computational complexity and ensured efficient convergence. However, the success of this strategy depends on interconnecting rules to avoid isolated groups that cannot be evaluated jointly.

Weights for interventions with non-zero contributions to specific nodes after optimization are presented in \textbf{Table 3}. Full weights are available upon request.

\textbf{Model Display}:  
To facilitate real-time testing and clinical usability, we developed a user interface allowing comparison of any two regimens (single or combination) at different time points. The interface is accessible at \url{https://namprotocols.org/predict.php}.

\begin{table}[ht]
\centering
\begin{tabular}{|l|p{10cm}|}
\hline
\textbf{Node}                          & \textbf{Intervention}                                                                                                                \\ \hline
Insulin Like growth factor (increased)  & Metformin, Spironolactone                                                                                                             \\ \hline
Free testosterone (increased)           & Metformin, Spironolactone, Oral contraceptive                                                                                        \\ \hline
Sebaceous gland activity                & Isotretinoin (standard dose), Isotretinoin (low dose)                                                                                \\ \hline
Keratinization (comedogenesis)          & Isotretinoin (standard dose), Isotretinoin (low dose)                                                                                \\ \hline
Comedone formation (follicular occlusion) & Adapalene, Tretinoin 0.025\%, Tretinoin 0.05\%, Tazarotene, Isotretinoin gel, Benzoyl peroxide, Azelaic acid, Azithromycin (pulse dosage), Roxithromycin \\ \hline
Bacterial growth (infection)            & Doxycycline, Azithromycin (pulse dosage), Lymecycline, Roxithromycin, Clindamycin, Benzoyl peroxide, Erythromycin, Dapsone, Ozenoxacin, Nadifloxacin, Azelaic acid, Tazarotene                                                               \\ \hline

Inflammation                            & Doxycycline, Azithromycin (pulse dosage), Lymecycline, Roxithromycin, Benzoyl peroxide, Erythromycin, Dapsone, Ozenoxacin, Nadifloxacin, Azelaic acid, Tazarotene, Adapalene, Tretinoin 0.025\%, Isotretinoin gel                                                                \\ \hline

\end{tabular}
\caption{Interventions having non-zero weight for each node after optimization}
\label{tab:non_zero_weight}
\end{table}

\subsection*{3.3 Prediction Accuracy Evaluation}

\subsection*{3.3.1 Model Evaluation on Previously Tested Regimens}

\textbf{Evaluation Dataset}  

We searched the CENTRAL database (Cochrane) for acne studies published between 2013 and August 2023, using predefined inclusion and exclusion criteria:

\begin{itemize}
    \item \textbf{Inclusion Criteria}:  
    \begin{itemize}
        \item Studies where both arms consisted of interventions from the selected list (\textbf{Table 1}).  
        \item Studies with unambiguous outcomes (a clear winner between regimens).  
    \end{itemize}
    \item \textbf{Exclusion Criteria}:  
    \begin{itemize}
        \item Poorly designed studies with inadequately presented results.  
        \item Comparisons where one arm included a regimen and the other arm was a combination of the same regimen with an additional treatment (e.g., adapalene vs. adapalene plus clindamycin), unless unexpected results were reported.
    \end{itemize}
\end{itemize}

Each study instance was defined by unique PICO parameters (i.e. patient population, intervention/comparison regimen, and outcome). Studies could generate multiple instances based on differences in arms, time intervals, or acne types. Instances with conflicting results or those without meaningful insights were excluded.

\textbf{Testing Procedure}  

For each evaluation instance, the two regimens were simulated in our system (we will therefore call it ECF to avoid confusion), and the predicted superior regimen was compared with trial outcomes. The graphical interface described earlier provided predictions on regimen efficacy based on node-level weight reductions. Prediction confidence levels were not considered in this analysis.

\textbf{Figure 2 (a \& b)} illustrates a hypothetical comparison between doxycycline plus adapalene and doxycycline plus azelaic acid using the ECF interface. All instances, the actual outcome and prediction by ECF is shown in \textbf{supplementary file}.

\textbf{Outcome}  

ECF achieved a prediction accuracy of 76\% (32 correct predictions out of 42 instances) against extracted comparison instances, with \textbf{Cohen’s kappa} between \textbf{ECF and clinical trial prediction being 0.50}, signifying \textbf{moderate to substantial agreement}, demonstrating its feasibility to align with clinical evidence for tested combinations. 

\begin{figure}[htp]
    \centering
    \begin{subfigure}[b]{0.45\textwidth}
        \centering
        \includegraphics[width=\textwidth]{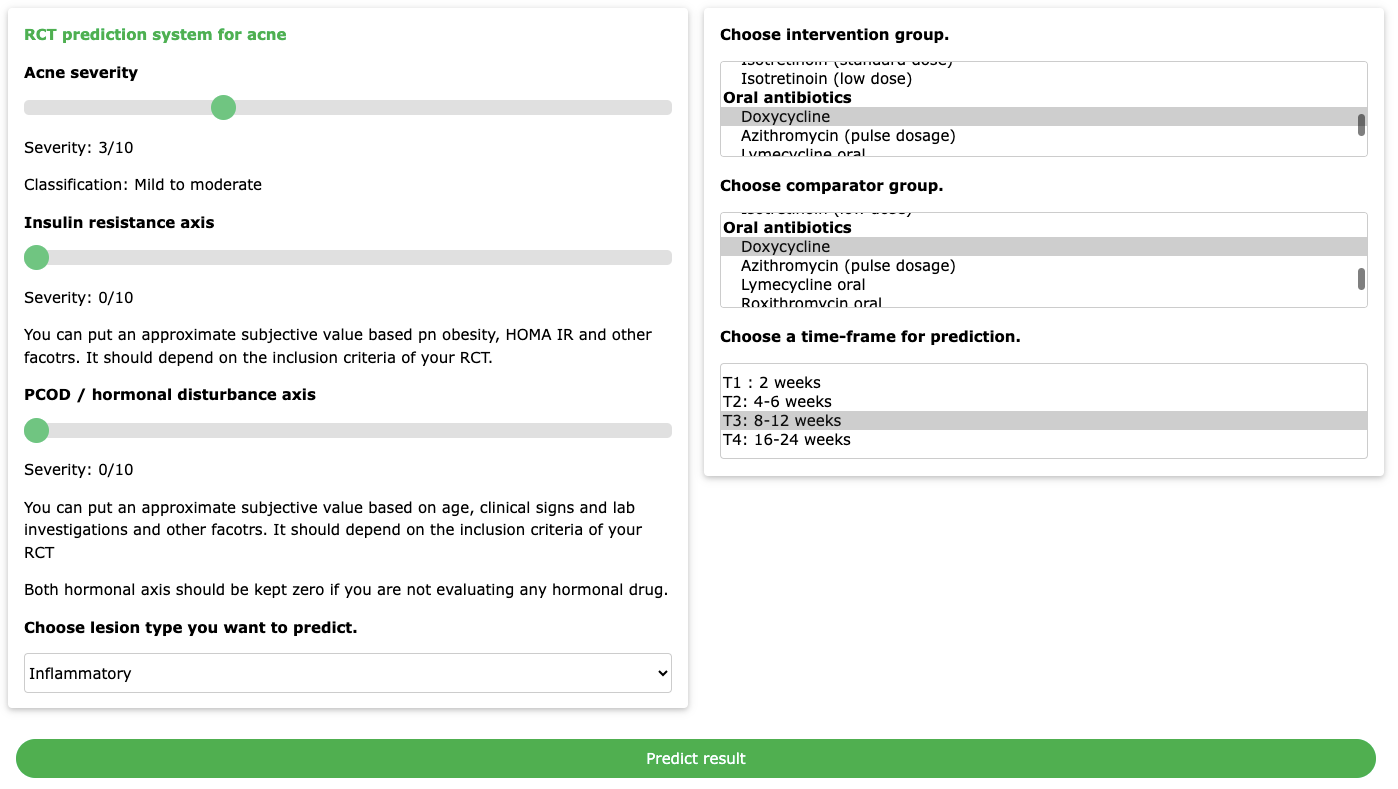}
        \caption{The user interface of trial prediction by ECF. In this imaginary trial, doxycycline and adapalene (intervention group; both selections not visible) are being compared with doxycycline and azelaic acid (comparator group; both selections not visible) for mild to moderate inflammatory acne vulgaris at 8-12 weeks.
 }
        \label{fig:2a}
    \end{subfigure}
    \hfill
    \begin{subfigure}[b]{0.45\textwidth}
        \centering
        \includegraphics[width=\textwidth]{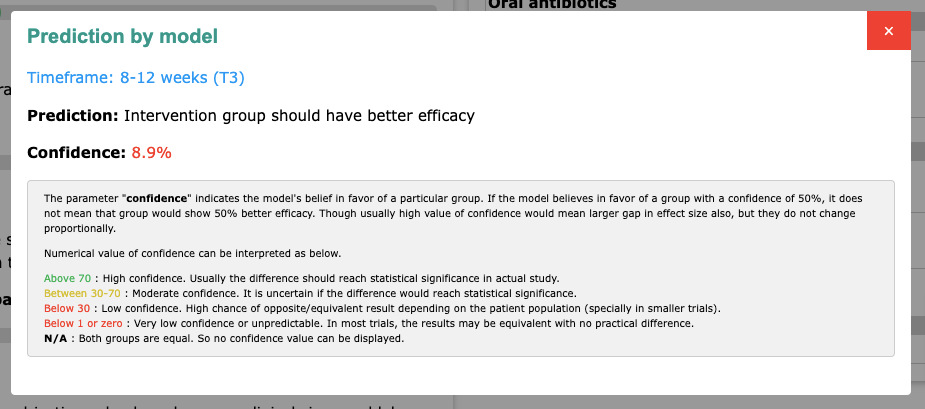}
        \caption{Prediction of the comparison by ECF. According to ECF, The intervention group (doxycycline and adapalene group) should have better efficacy by 12 weeks. However, the confidence is less than 30 (8.9\% only) indicating very low confidence.
}
        \label{fig:2b}
    \end{subfigure}
    \caption{}
    \label{fig:main}
\end{figure}

\subsection*{3.3.2 Predicting Outcomes for Previously Untested Regimens}

To further assess the generalizability of ECF beyond validated datasets, we conducted an independent evaluation using \textbf{previously untested regimen comparisons}, benchmarked against \textbf{expert dermatologist assessments}.

\subsubsection*{Study Design}
\begin{itemize}
    \item \textbf{Generation of Comparisons}: 50 potential regimen comparisons were randomly generated under clinical constraints (time frame, acne type, treatment category).
    \item \textbf{Filtering Process}: 26 rational and feasible regimen comparisons were selected after excluding impractical combinations.
    \item \textbf{Expert Dermatologist Evaluation}: Three board-certified dermatologists (D1, D2, D3; experience: 15–50 years) provided \textbf{blinded independent predictions}, based solely on efficacy intuition. None of these dermatologists had any previous or concurrent access to our model prior to or during  the evaluation process.
    \item \textbf{Consensus Benchmarking}: A majority-vote consensus (C) was established for each comparison (i.e., \(\geq2\) out of 3 agreement).
\end{itemize}

\subsubsection*{Agreement Analysis}
To determine the degree of alignment between ECF and expert predictions, \textbf{Cohen’s kappa (\(\kappa\))} was computed for each comparison:

\begin{itemize}
    \item \textbf{ECF vs. Individual Dermatologists:}
    \begin{itemize}
        \item \textbf{D1}: \(\kappa = 0.49\)
        \item \textbf{D2}: \(\kappa = 0.43\)
        \item \textbf{D3}: \(\kappa = 0.37\)
        \item \textbf{Consensus (C)}: \(\kappa = 0.49\)
    \end{itemize}
    \item \textbf{Inter-dermatologist Agreement:}
    \begin{itemize}
        \item \textbf{D1 vs. D2}: \(\kappa = 0.44\)
        \item \textbf{D2 vs. D3}: \(\kappa = 0.61\)
        \item \textbf{D1 vs. D3}: \(\kappa = 0.22\)
        \item \textbf{Overall Fleiss’ kappa among all three}: \(\kappa = 0.42\)
    \end{itemize}
\end{itemize}

\subsubsection*{Comparative Performance: ECF on Tested vs. Untested Regimens}
\begin{itemize}
    \item ECF’s \textbf{accuracy on untested regimens} (vs. dermatologist consensus) was \textbf{76\% (20/26 correct predictions)}and  \textbf{Cohen’s kappa for ECF vs. Consensus}: \(\kappa = 0.49\).  Comparing this with tested regimen (76\%,\(\kappa = 0.50\)) , there was \textbf{no significant difference (p = 0.977)} by Z test. The result is shown in \textbf{Figure 3}. The moderate inter-expert agreement (Fleiss’ \(\kappa\) = 0.42) underscores the challenge of standardizing efficacy assessments, even among specialists.
\end{itemize}

\begin{figure}
    \centering
    \includegraphics[width=0.55\linewidth]{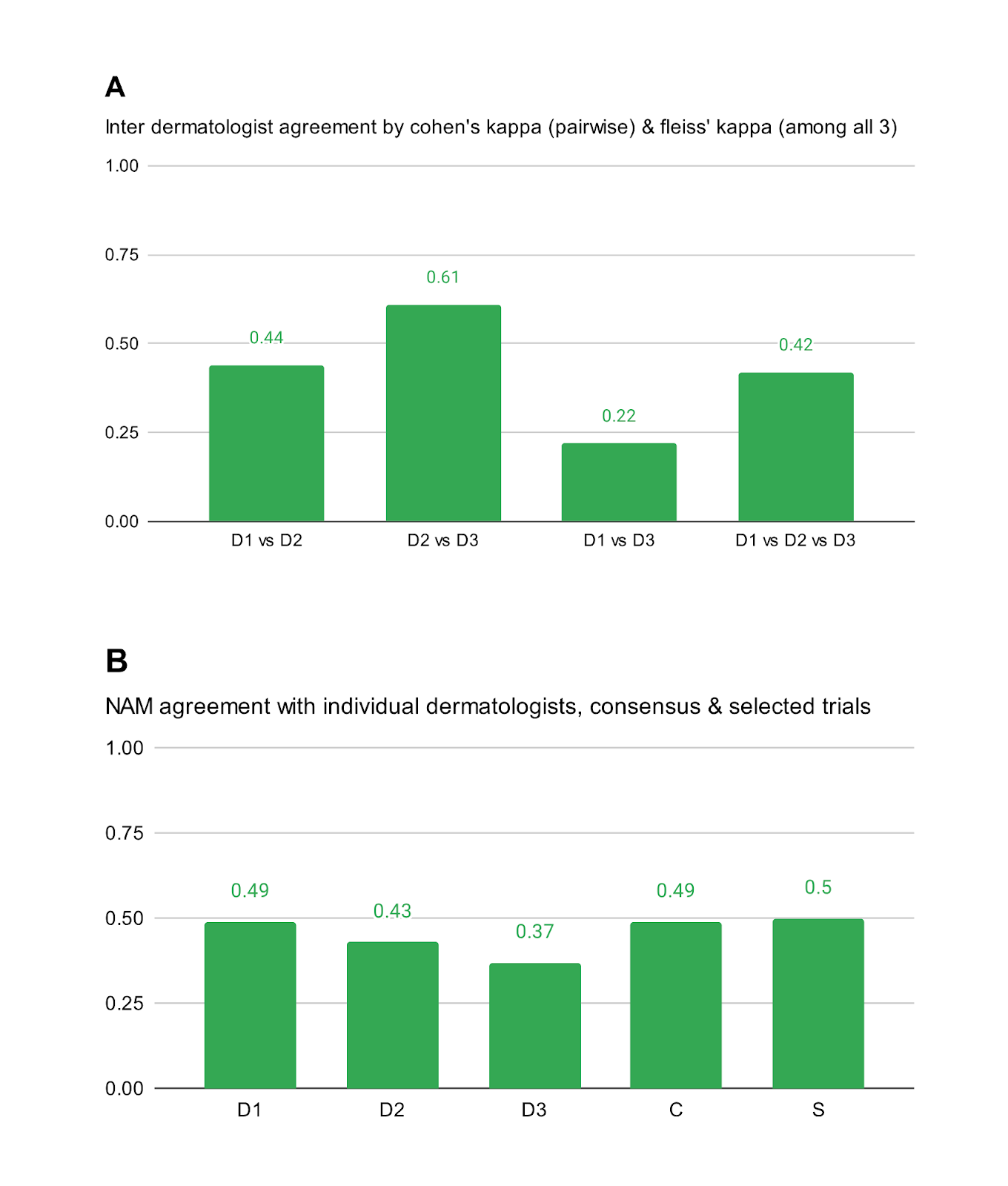}
    \caption{(A) Inter-rator agreement among dermatologists (D1, D2 and D3), pairwise comparison done by kohen’s kappa and fleiss’ kappa for all 3. (B) Agreement between prediction by ECF-based Acne model and individual dermatologists, their consensus (C), and selected study results (S).}
    \label{fig:enter-label}
\end{figure}

\subsection*{3.4 Statistical and Clinical Significance}

ECF’s predictive accuracy \textbf{remains stable across tested and untested regimens}, \textbf{indicating no over-fitting}. The agreement scores between \textbf{ECF and dermatologists are statistically comparable to inter-dermatologist agreement}, suggesting that the model \textbf{mirrors expert clinical intuition}.

\section*{4. Discussion}

The \textbf{Efficacy Comparison framework (ECF)} is a novel, mechanistically grounded system for predicting the efficacy of combination therapies by integrating domain-specific biological insights and clinical evidence. ECF offers a solution to several persistent challenges in evidence synthesis, such as parameter explosion, limited interpretability, and the incorporation of domain knowledge, while maintaining scalability and computational efficiency. 

\subsection*{4.1 ECF and the Landscape of Synergy Prediction}

The current paradigm of synergy prediction relies heavily on high-throughput screening [31] or multi-omics-driven machine learning models (e.g., DeepSynergy [32], AuDNNsynergy [33]), which are feasible only in data-rich contexts such as oncology, where cell lines and multi-omics datasets enable complex model optimization. For most diseases—particularly those where in vitro simulations are infeasible (e.g., chronic inflammatory or multifactorial disorders)—clinicians must rely solely on sparse clinical trial data, limiting the applicability of data-hungry methods. Traditional statistical frameworks like component network meta-analysis (cNMA) face fundamental trade-offs: additive models oversimplify interactions [4], while interaction models suffer from combinatorial parameter explosions. ECF addresses this gap by harmonizing model complexity with available evidence. By encoding domain knowledge into a mechanistic graph structure, ECF avoids reliance on chemical similarity biases [31] or protein-protein interaction (PPI) topologies [5], which often ignore disease-specific pathway hierarchies [34]. Instead, ECF dynamically adjusts its parameter space (as discussed in \textbf{Section 1.4.3}) to reflect the scale of clinical trial data, enabling synergy predictions grounded in biological plausibility rather than purely statistical assumptions. This positions ECF as a scalable intermediary—less data-intensive than multi-omics models yet more interpretable and mechanistically robust than cNMA—making it uniquely suited for diseases where clinical evidence is limited but domain expertise is abundant.

\subsection*{4.2 Alignment of ECF with Mechanistic Principles}

ECF is designed as a generalizable framework that integrates domain-specific principles (e.g., complementary targeting [5]) through configurable components such as aggregator functions. For instance, selecting Bliss Independence inherently encodes synergy for non-overlapping pathways, aligning with network science principles without requiring explicit PPI topology [34]. Crucially, ECF’s modularity allows incorporation of diverse mechanistic evidence (e.g., pathway hierarchies, pharmacokinetic interactions) as data availability permits, scaling complexity via its graph structure (Section 1.4.3). By prioritizing disease pathogenesis graphs over generic interactomes, ECF bridges clinical relevance and computational flexibility: its parameters reflect actionable biological insights rather than abstract network properties, while remaining adaptable to new evidence or therapeutic paradigms.

\subsection*{4.3 Implications for Clinical Guideline Development}

Current clinical guideline development lacks a standardized framework for integrating mechanistic insights with empirical evidence [35]. While existing guidelines layer expert consensus atop trial data, ECF embeds domain knowledge directly into the evidence synthesis process, enabling a more systematic and scalable approach. By encoding biological pathways and drug mechanisms into its graph structure, ECF provides a transparent, interpretable foundation for dynamic treatment recommendations. For instance, patient-specific factors (e.g., genetic mutations, comorbidities) can be incorporated as additional nodes, adjusting a node's value or additional constraints, allowing tailored regimen predictions for individual patients or subgroups. Furthermore, by integrating side-effects and toxicity data as optimization constraints, ECF can evolve into a comprehensive therapy optimization framework. This adaptability ensures scalability: as new evidence emerges, the graph can be updated to reflect evolving mechanistic understanding or clinical priorities. In this way, ECF bridges the gap between traditional guidelines—which rely on static, population-level evidence—and the future of personalized medicine, where treatment decisions are dynamically informed by both mechanistic insights and patient-specific data.

\subsection*{4.4 Limitations and Future Directions}

\textbf{Integration with Advanced Network Science} :
As ECF incorporates mechanistic graphs, future iterations could benefit from deeper integration with network science insights depending on data availability. For example:
\begin{itemize}
    \item Incorporating local and global topological features from PPI networks to refine node weights or connectivity.
    \item Leveraging multi-omics data to construct more robust pathogenesis graph [\textit{36}] and dynamically update as new evidence emerges.
\end{itemize}

\textbf{Standardization of Parameters and Rules} :
Currently, our framework relies on expert-derived  quasi-rules, which may introduce subjectivity into the model. Future work should focus on automating this process through standardized, data-driven approaches, such as natural language processing (NLP) or Large Language Model (LLM) for extracting rules from literature or integrating large-scale datasets to inform initial parameter values.

\textbf{Addressing Toxicity and Adverse Effects}: 
While ECF focuses on efficacy, future extensions could incorporate toxicity predictions, aligning with the principle that synergistic drugs should minimize overlapping toxicities. This would require expanding the mechanistic graph to include nodes representing adverse outcomes or resource competition at the cellular level. This is important because without this constraint, therapy optimization is not possible.

\section*{Conclusion}
The Efficacy Comparison Framework (ECF) offers a general system for comparing combination therapies (including previously untested ones), bridging the gap between clinical trial data and mechanistic insights through scalable modeling. By incorporating biological pathways as mechanistic abstractions, ECF effectively addresses critical challenges such as parameter explosion, limited comparison availability, and the difficulty of encoding domain knowledge into the model. Integrating expert-derived priors and clinical trial data into a biologically interpretable structure ensures relevance and applicability across various therapeutic domains. By design, ECF integrates domain knowledge to compensate for sparse clinical trial data, offering a scalable alternative to additive cNMA and interaction cNMA.

ECF's flexible alignment with key principles from network science and synergy prediction models highlights its robustness. The use of disease pathogenesis graphs as an abstraction layer provides a flexible, scalable alternative to models relying solely on PPI networks or chemical similarity, while the choice of effect-based aggregator functions ensures compatibility with clinical trial data. The approach demonstrated promising predictive accuracy in a case study of acne vulgaris, showing its potential to guide treatment decisions even for untested regimens.

However, ECF is not without limitations. Standardization of parameterization and incorporation of toxicity data remain areas for further enhancing its scope. Future iterations of ECF could also benefit from deeper integration with multi-omics data, adaptive learning systems, and advanced network science methodologies.

In conclusion, ECF represents a scalable, interpretable, and biologically grounded approach to combination therapy prediction. It provides a foundation for future developments in evidence synthesis, personalized medicine, and guideline development, ensuring its utility in clinical decision-making and advancing the field of systems pharmacology.

\section*{Acknowledgments}
We are personally indebted to the three senior academic dermatologists, Prof Debabrata Bandyopadhyay, Dr Sandipan Dhar and Dr Anupam Das, for agreeing unconditionally to take part in the experimental process of independent blinded prediction, based solely on efficacy intuition, that is a critical part of this study.

\section*{Competing Interest}

The authors declare that they have no competing interests or personal relationships that could have influenced the work reported in the article.

\section*{Funding sources}

This research did not receive any grant from any funding agencies.

\section*{Data Sharing Statement}

Upon reasonable request, we can share the full post optimization weights of our ECF-based acne model. The intended use must be academic \& non-commercial in nature.

\section*{Ethics Approval}

No Ethics Approval (Not applicable)

\section*{Declaration of generative AI and AI-assisted technologies in the writing process}
During the preparation of this work the authors used GPT-4o (ChatGPT) to improve readability and minor language formatting. After using this tool/service, the authors reviewed and edited the content as needed and takes full responsibility for the content of the publication.

\section*{References}

\begin{enumerate}
    \item He B, Lu C, Zheng G, He X, Wang M, Chen G, Zhang G, Lu A. Combination therapeutics in complex diseases. J Cell Mol Med. 2016 Dec;20(12):2231-40.
    \item Welton NJ, Caldwell DM, Adamopoulos E, Vedhara K. Mixed treatment comparison meta-analysis of complex interventions: psychological interventions in coronary heart disease. Am J Epidemiol. 2009 May 1;169(9):1158-65.
    \item Rouse B, Chaimani A, Li T. Network meta-analysis: an introduction for clinicians. Intern Emerg Med. 2017 Feb;12:103-11.
    \item R\"ucker G, Petropoulou M, Schwarzer G. Network meta‐analysis of multicomponent interventions. Biometrical J. 2020 May;62(3):808-21.
    \item Cheng F, Kov\'acs IA, Barab\'asi AL. Network-based prediction of drug combinations. Nat Commun. 2019 Mar 13;10(1):1197.
    \item Bradley RA, Terry ME. Rank analysis of incomplete block designs: I. The method of paired comparisons. Biometrika. 1952 Dec 1;39(3-4):324-45.
    \item Han R, Ye R, Tan C, Chen K. Asymptotic theory of sparse Bradley–Terry model. Ann Appl Probab. 2020 Oct;30(5):2491-515.
    \item Anderson PW. More is different: broken symmetry and the nature of the hierarchical structure of science. Science. 1972 Aug 4;177(4047):393-6.
    \item Jamieson KG, Nowak RD. Low-dimensional embedding using adaptively selected ordinal data. In: 49th Annual Allerton Conference on Communication, Control, and Computing (Allerton); 2011 Sep 28. p. 1077-84. IEEE.
    \item Candes EJ, Recht B. Exact low-rank matrix completion via convex optimization. In: 46th Annual Allerton Conference on Communication, Control, and Computing; 2008 Sep 23. p. 806-12. IEEE.
    \item Springall A. Response surface fitting using a generalization of the Bradley-Terry paired comparison model. J R Stat Soc C Appl Stat. 1973 Mar;22(1):59-68.
    \item De Soete G, Winsberg S. A Thurstonian pairwise choice model with univariate and multivariate spline transformations. Psychometrika. 1993 Jun;58(2):233-56.
    \item Foucquier J, Guedj M. Analysis of drug combinations: current methodological landscape. Pharmacol Res Perspect. 2015 Jun;3(3):e00149.
    \item Cover TM. Elements of information theory. New York: John Wiley \& Sons; 1999.
    \item Murphy KP. Dynamic Bayesian networks: representation, inference and learning. Berkeley: University of California; 2002.
    \item Brookes DH, Otwinowski J, Sinai S. Contrastive losses as generalized models of global epistasis. arXiv preprint arXiv:2305.03136. 2023 May 4.
    \item Thurstone LL. A law of comparative judgment. In: Scaling. Routledge; 2017 Jul 5. p. 81-92.
    \item Cattelan M. Models for paired comparison data: a review with emphasis on dependent data.
    \item Zaenglein AL, Pathy AL, Schlosser BJ, Alikhan A, Baldwin HE, Berson DS, et al. Guidelines of care for the management of acne vulgaris. J Am Acad Dermatol. 2016 May;74(5):945-73.
    \item National Institute for Health and Care Excellence. Acne vulgaris: management [Internet]. Available from: \url{https://www.nice.org.uk/guidance/ng198}.
    \item Oon HH, Wong SN, Aw DC, Cheong WK, Goh CL, Tan HH. Acne management guidelines by the Dermatological Society of Singapore. J Clin Aesthet Dermatol. 2019 Jul;12(7):34.
    \item Hayashi N, Akamatsu H, Iwatsuki K, Shimada‐Omori R, Kaminaka C, Kurokawa I, et al. Japanese Dermatological Association Guidelines: Guidelines for the treatment of acne vulgaris 2017. J Dermatol. 2018 Aug;45(8):898-935.
    \item Bruinsma M, Jaspar A, De Ruijter W, et al. NHG-werkgroep acne. NHG-Standaard Acne (derde herziening). Huisarts Wet. 2017;4:164-70.
    \item Le Cleach L, Lebrun‐Vignes B, Bachelot A, Beer F, Berger P, Brug\`ere S, et al. Guidelines for the management of acne: recommendations from a French multidisciplinary group. Br J Dermatol. 2017 Oct 1;177(4):908-13.
    \item Thiboutot DM, Dr\'eno B, Abanmi A, Alexis AF, Araviiskaia E, Cabal MI, et al. Practical management of acne for clinicians: an international consensus from the Global Alliance to Improve Outcomes in Acne. J Am Acad Dermatol. 2018 Feb 1;78(2 Suppl 1):S1-23.
    \item Nast A, Dr\'eno B, Bettoli V, Bukvic Mokos Z, Degitz K, Dressler C, et al. European evidence‐based (S3) guideline for the treatment of acne–update 2016–short version. J Eur Acad Dermatol Venereol. 2016 Aug;30(8):1261-8.
    \item Brouwers MC, Kho ME, Browman GP, Burgers JS, Cluzeau F, Feder G, et al. AGREE II: advancing guideline development, reporting and evaluation in health care. CMAJ. 2010 Dec 14;182(18):E839-42.
    \item Berenbaum MC. What is synergy?. Pharmacol Rev. 1989 Jun 1;41(2):93-141.
    \item Bliss CI. The calculation of microbial assays. Bacteriol Rev. 1956 Dec;20(4):243-58.
    \item He L, Kulesskiy E, Saarela J, et al. Methods for high-throughput drug combination screening and synergy scoring. In: von Stechow L, editor. Cancer Systems Biology. Clifton, NJ: Springer; 2018. p. 351-98.
    \item Wu L, Wen Y, Leng D, Zhang Q, Dai C, Wang Z, et al. Machine learning methods, databases and tools for drug combination prediction. Brief Bioinform. 2022 Jan;23(1):bbab355.
    \item Preuer K, Lewis RP, Hochreiter S, Bender A, Bulusu KC, Klambauer G. DeepSynergy: predicting anticancer drug synergy with deep learning. Bioinformatics. 2018;34(9):1538-46.
    \item Zhang T, Zhang L, Payne PR, Li F. Synergistic drug combination prediction by integrating multiomics data in deep learning models. In: Walker J, editor. Translational Bioinformatics for Therapeutic Development. New York, NY: Springer; 2021. p. 223-38.
    \item Li H, Li T, Quang D, Guan Y. Network propagation predicts drug synergy in cancers. Cancer Res. 2018;78(18):5446-57.
    \item Djulbegovic B, Hozo I, Lizarraga D, Guyatt G. Decomposing clinical practice guidelines panels' deliberation into decision theoretical constructs. J Eval Clin Pract. 2023 Apr;29(3):459-71.
    \item Viswanathan GA, Seto J, Patil S, Nudelman G, Sealfon SC. Getting started in biological pathway construction and analysis. PLoS Comput Biol. 2008 Feb;4(2):e16.

\end{enumerate}

\end{document}